\documentclass{nature}

\usepackage{upgreek}
\usepackage{amssymb}
\usepackage{times}
\usepackage{graphicx}
\usepackage{bm}

\newcommand{\CII}{[C\,\textsc{ii}]}

\newcommand{\HI}{{H\,\textsc{i}}}

\newcommand{\kms}{km~s$^{-1}$}

\makeatletter
\let\saved@includegraphics\includegraphics
\AtBeginDocument{\let\includegraphics\saved@includegraphics}
\renewenvironment*{figure}{\@float{figure}}{\end@float}
\makeatother

\title{A Cold, Massive, Rotating Disk 1.5 Billion Years after the Big Bang}

\author{Marcel Neeleman$^{1}$, J. Xavier Prochaska$^{2,3}$,
Nissim Kanekar$^{4}$ \& Marc Rafelski$^{5,6}$}

\begin{document}

\maketitle

\begin{affiliations}
\item Max-Planck-Institut f\"{u}r Astronomie, K\"{o}nigstuhl 17, D-69117, Heidelberg, Germany
\item Department of Astronomy \& Astrophysics, UCO/Lick Observatory, University of California, 1156 High Street, Santa Cruz, CA 95064, USA
\item Kavli Institute for the Physics and Mathematics of the Universe (Kavli IPMU), The University of Tokyo, 5-1-5 Kashiwanoha, Kashiwa, 277-8583, Japan
\item National Centre for Radio Astrophysics, Tata Institute of Fundamental Research, Pune University, Pune 411007, India
\item Space Telescope Science Institute, Baltimore, MD 21218, USA
\item Department of Physics \& Astronomy, Johns Hopkins University, Baltimore, MD 21218, USA
\end{affiliations}

\begin{abstract}
Massive disk galaxies like the Milky Way are expected to form at late times in traditional models of galaxy formation\cite{Rees1977,Fall1980}, but recent numerical simulations suggest that such galaxies could form as early as a billion years after the Big Bang through the accretion of cold material and mergers\cite{Keres2005,Dekel2009}. Observationally, it has been difficult to identify disk galaxies in emission at high redshift\cite{Hodge2012,Smit2018}, in order to discern between competing models of galaxy formation. Here we report imaging, with a resolution of about 1.3 kiloparsecs, of the 158-micrometre emission line from singly ionized carbon, the far-infrared dust continuum and the near-ultraviolet continuum emission from a galaxy at a redshift of 4.2603, identified by detecting its absorption of quasar light. These observations show that the emission arises from gas inside a cold, dusty, rotating disk with a rotational velocity of 272 kilometres per second. The detection of emission from carbon monoxide in the galaxy yields a molecular mass that is consistent with the estimate from the ionized carbon emission of about 72 billion solar masses. The existence of such a massive, rotationally supported, cold disk galaxy when the Universe was only 1.5 billion years old  favours formation through either cold-mode accretion or mergers, although its large rotational velocity and large content of cold gas remain challenging to reproduce with most numerical simulations\cite{Grand2017,Pillepich2019}.
\end{abstract}

An open question in galaxy evolution is the epoch at which disk galaxies like our Milky Way formed. In our current cosmology paradigm, known as $\Lambda$-Cold Dark Matter\cite{Planck2016}, galaxies are expected to be built up in a hierarchical manner. Gas and dark matter funnel into dark matter halos, merging and condensing into larger structures, precipitating the formation of stars and the growth of the galaxy. However, the physical processes that dominate galaxy formation are still under debate. 

In the traditional picture of galaxy formation, the infalling gas is shock-heated to the virial temperature (about 10$^6$~K for a 10$^{12}$ solar mass ($M_\odot$) galaxy) and accretes spherically; the central region then cools and condenses into a rotationally-supported disk\cite{Rees1977,Fall1980}. Besides this `hot mode' accretion scenario, numerical simulations predict an alternative scenario in which gas accretes efficiently onto galaxies either through the merging of galaxies or through gas flowing directly into galaxies along filamentary structures, with a considerable fraction of the gas remaining cool, at temperatures far below the virial temperature of the galaxies\cite{Keres2005,Dekel2009}. Unlike the hot-mode scenario, in which the long cooling times imply that disk galaxies form at relatively late times (redshift $z \lesssim$ 1), in these latter scenarios, disk galaxies can form much earlier ($z \lesssim 5$)\cite{Grand2017,Pillepich2019}. Observing the earliest onset of galaxy disks can therefore inform us how galaxies acquire their mass, allowing us to distinguish between these mass accretion scenarios.

Observationally, disks have been identified via carbon monoxide (CO) or H$\upalpha$ spectroscopy at $z \lesssim$ 2.5 (refs. \cite{Forster-Schreiber2009,Price2016,Genzel2017}). At higher redshifts, $z \approx 4-5$, observations with the Karl G. Jansky Very Large Array (JVLA) and Atacama Large Millimetre/submillimetre Array (ALMA) have yielded tentative evidence of disks\cite{Hodge2012,Smit2018}. However, a combination of relatively low resolution and sensitivity has meant that it has been impossible to conclusively identify rotating disk galaxies at $z \gtrsim 3$. The vast majority of these studies have focused on objects with high star-formation rates (SFR), selected by their high luminosity in either the optical/near-infrared and/or at sub-millimetre wavelengths. Because the star formation efficiency is higher in dense, clumpy environments, such `emission-selected' galaxy samples could be biased against the presence of stable disks. A complementary approach to identifying high-redshift galaxies lies in detecting absorption lines from gas in the galaxy, if the galaxy happens to lie in front of a bright background source such as a quasi-stellar object (QSO). This approach does not contain a bias towards more luminous galaxies, and therefore `absorption-selected' galaxy samples provide a unique luminosity-unbiased sample to understand disk galaxy formation at high redshifts.

Our recent ALMA searches for absorption-selected galaxies using the fine-structure line of singly ionized carbon at 157.74 $\upmu$m (\CII) have revealed a sample of six galaxies at $z$~$\approx$~4 (refs.~\cite{Neeleman2017,Neeleman2019}), with SFR of  about 7 $-$ 110 $M_\odot$~yr$^{-1}$, as determined from their \CII\ and dust continuum emission. Owing to the coarse angular resolution of these observations ($\approx$1$''$, which corresponds to about 6.5 kpc at the redshift of the galaxies), we were unable to characterize the dynamics of the gas inside the galaxies. However, half of the galaxies showed a tentative velocity gradient in their \CII\ emission\cite{Neeleman2017,Neeleman2019}, consistent with the \CII\ arising from a rotating disk. To explore the origin of this \CII\ emission, we carried out high-resolution (about 0.19$''$, corresponding to approximately 1.3 kpc at the redshift of the galaxy) ALMA observations of the \CII\ and dust continuum emission from the brightest \CII-emitting galaxy of this sample, DLA0817g, which is associated with a $z~=$~4.26 absorber towards QSO~J081740.52+135134.5.

The \CII\ emission from DLA0817g is resolved in the imaging at a resolution of 0.19$''$ (1.3 kpc), and has an extent of 0.63~$\pm$~0.07$''$ (4.2~$\pm$~0.5~kpc), along the major axis of the galaxy (see Methods). Within this extent, the velocity gradient of the \CII\ line remains smooth even at 0.19$''$ resolution (Fig.~1a), indicating that the \CII\ emission arises from a rotationally supported disk. This is further corroborated by the position-velocity ($p$-$v$) diagram of DLA0817g (Fig.~1b), which is a slice of the spectral cube, along the galaxy's major axis. The $p$-$v$ diagram shows a flattening of the velocity curve at about 1.8~kpc, indicating that the gas has reached a constant rotational velocity. This characteristic S-shape is the proto-typical signature of a rotating disk\cite{DeBlok2008}. 

Using a custom, python-based Markov Chain Monte Carlo code, we have fitted the observations with a rotating disk model (Methods). The best-fit model has a position angle of 105.2~$\pm$~1.7$^{\rm o}$, an inclination angle of 42$^{+3}_{-8}$$^{\rm o}$ and an inclination-corrected rotational velocity of $v_{\rm rot}$ = 272$^{+52}_{-13}$~\kms. The residuals after subtracting the disk model are below 3$\sigma$ significance, where $\sigma$ is the standard deviation of the noise, and show little velocity offset, indicating that the bulk of the \CII\ emission can be modelled as a rotating disk (Fig.~2). With this position angle and inclination, we reconstruct the rotation curve of DLA0817g (Fig.~3) through two different approaches (peak velocity and mean velocity; see Methods). The rotation curve is flat beyond a radius of about 1.8~kpc, with a mean rotational velocity that is consistent with the value from the kinematic modelling. The observed decrease in the rotation curve below this radius is due to resolution effects (also known as beam-smearing); the rotating disk model, which has a constant velocity, shows the same decrease when convolved to the resolution of the data. Combining the rotational velocity estimate with the maximum extent of the \CII\ emission yields a dynamical mass estimate of (7.2~$\pm$~2.3)~$\times$~10$^{10}$~$M_\odot$ (Methods).

Having established the disk origin of the \CII\ emission, we can provide an estimate of the rotational support and the stability of the gas against axisymmetric perturbations. The ratio of rotational velocity to velocity dispersion ($v_{\rm rot}/\sigma_v$) provides a measure of the rotational support of a galaxy, with ratios greater than 3 indicating a galaxy that is rotation-dominated\cite{Burkert2010}. For DLA0817g, the estimates of both the rotational velocity and the velocity dispersion come from the kinematic modelling of the \CII\ emission line. These estimates are consistent with measurements of the rotational velocity and velocity dispersion away from the centre of emission where beam-smearing is less severe (Methods). Our estimate for $v_{\rm rot}/\sigma_v$ is 3.4$^{+1.1}_{-0.3}$, consistent with DLA0817g being a rotation-dominated system. In addition to $v_{\rm rot}/\sigma_v$, the Toomre-$Q$ parameter of a disk provides a measure of the stability of the disk against gravitational fragmentation, where values of order unity indicate stable disks\cite{Toomre1964,Goldreich1965}. For DLA0817g, we obtain a disk-averaged Toomre-$Q$ parameter of 0.96 $\pm$ 0.30, close to the stability limit, which is consistent with predictions from theory\cite{Burkert2010}. However, local values of $Q$ within the disk can fall well below unity, resulting in unstable regions that should collapse to form dense gas, and then stars\cite{Elmegreen2008}. Such dense gas is expected to show weaker \CII\ emission, as most of the carbon is locked up in CO (ref.~\cite{Beuther2014}), and the increase in dust will attenuate the \CII\ emission\cite{Riechers2013}. The \CII\ cavity in DLA0817g, about 2~kpc east of the galaxy centre (Fig.~1), may arise from such a locally unstable region.

Although \CII\ emission is a good tracer of the dynamics of a galaxy\cite{DeBlok2016}, it can originate from gas with a wide range of physical properties\cite{Croxall2017,Cormier2019}. Here we examine how the resolved \CII\ observations compare to tracers of different mass constituents of the galaxy (that is, the dust, stellar, and molecular gas components). The underlying far-infrared continuum emission in the ALMA observations arises from reprocessed dust, and thus traces the spatial distribution of dust\cite{Carilli2013}. To trace the stellar properties of the galaxy, we observed DLA0817g with the Wide Field Camera 3 on the Hubble Space Telescope (HST), using the F160W filter. These observations cover the rest-frame near-ultraviolet (UV; about 300~nm) emission from the galaxy at an angular resolution of 0.31$''$. Overlaying the \CII\ and dust continuum contours on the HST image (Fig.~4b) shows that the emission centroids of all the tracers agree within the uncertainties, as do their effective radii (Methods). The dust continuum emission shows no evidence of substructure on scales of approximately 1.3~kpc, consistent with an origin in a smooth disk. The stellar emission from DLA0817g is extended with a physical size of 1.2~$\pm$~0.5$''$ (8.1~$\pm$~3.4~kpc) along the major axis of the galaxy at a position angle of 114~$\pm$~8$^{\rm o}$, within one standard deviation from the position angle obtained from the ALMA observations. The similar extent and shape of the near-UV, \CII\ and dust emission suggests that the \CII\ emission predominantly traces gas that is co-spatial with the stars and the dust.

To estimate the molecular content of DLA0817g, we observed the redshifted CO(2$-$1) transition with the JVLA. The CO rotational lines provide the best tracer of the molecular gas, as CO is the second-most abundant molecule (after molecular hydrogen, which is difficult to detect directly) in the interstellar medium of galaxies. The JVLA observations yield a detection of the CO(2$-$1) line at the position of DLA0817g (Fig.~4a), which results in an estimate of the molecular mass of (8.8~$\pm$~2.6)~$\times$~10$^{10}$~$\times$~(0.81 / r$_{\,2,1}$)~$\times$~($\alpha_{\rm CO}$ / 3.0)~$M_\odot$ (Methods). This molecular mass estimate is comparable with our estimate for the dynamical mass of DLA0817g, with the caveat that the unresolved JVLA observations can only probe the total molecular mass within the beam of the JVLA observations ($\sim$15~kpc). However, under the assumption that most of the molecular gas is constrained within the region of the galaxy that contains stars, dust and \CII\ emission\cite{Gullberg2018}, this implies that a substantial fraction of the galaxy's mass must reside in a cold, dense gas phase. 

DLA0817g was identified owing to an enriched neutral hydrogen (\HI) absorber at a projected distance of 6.2$''$ (42~kpc) from the galaxy\cite{Neeleman2017}, with an \HI\ column density of 2~$\times$~10$^{21}$~cm$^{-2}$. In the local Universe, such high \HI\ column densities are found only within the disks of galaxies\cite{Jones2018}. It is, however, unlikely that the absorption arises from an extension of the galaxy disk that we have imaged in \CII, as the implied disk radius of $\gtrsim 42$~kpc is far larger than values predicted by numerical  simulations\cite{Pillepich2019}. The absorbing gas is more likely to arise in a gas-rich clump inside an extended \HI\ reservoir that has been previously enriched by DLA0817g and is co-rotating with the disk\cite{Neeleman2017}. This further disfavours hot-mode accretion as the primary mass accretion scenario, as such high column densities of cold neutral gas are not expected so far from the central galaxy in this scenario.

The properties of DLA0817g and its associated absorber are typical for the full sample of metal-enriched, absorption-selected galaxies at these redshifts\cite{Neeleman2019}. Together with the selection method, through absorption, this suggests that such systems are common among normal star-forming galaxies at these redshifts. These observations therefore disfavour hot-mode accretion as the primary mass accretion method for these galaxies and support the existence of cold, rotationally-supported disk galaxies when the Universe was about 10\% of its current age.

\newpage
\begin{figure}[!t]
\includegraphics[width=\textwidth]{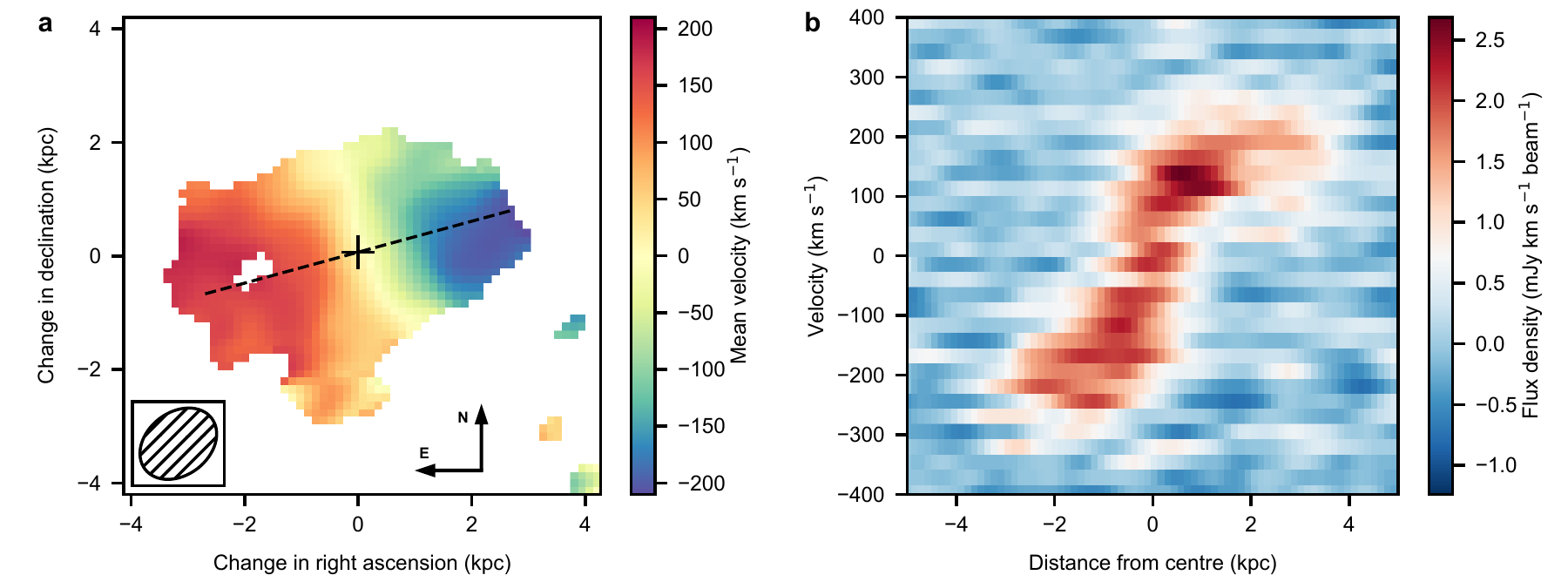}
\end{figure}
\noindent {\bf Fig. 1. Mean velocity field and $\bm p$-$\bm v$ diagram for DLA0817g.} {\bf a,} Mean velocity field relative to the systemic redshift of the \CII\ emission ($z =$~4.2603). The kinematic centre of the \CII\ emission, as determined from modelling the emission (see Methods), is shown by a black plus sign. The dotted black line marks the major axis of the galaxy. The axes give relative physical (proper) distances at the redshift of the \CII\ emission. The inset shows the synthesized beam of the observations. {\bf b,} The $p$-$v$ diagram along the major axis of the galaxy. Distances are measured from the kinematic centre of the galaxy.

\newpage
\begin{figure}[!t]
\includegraphics[width=\textwidth]{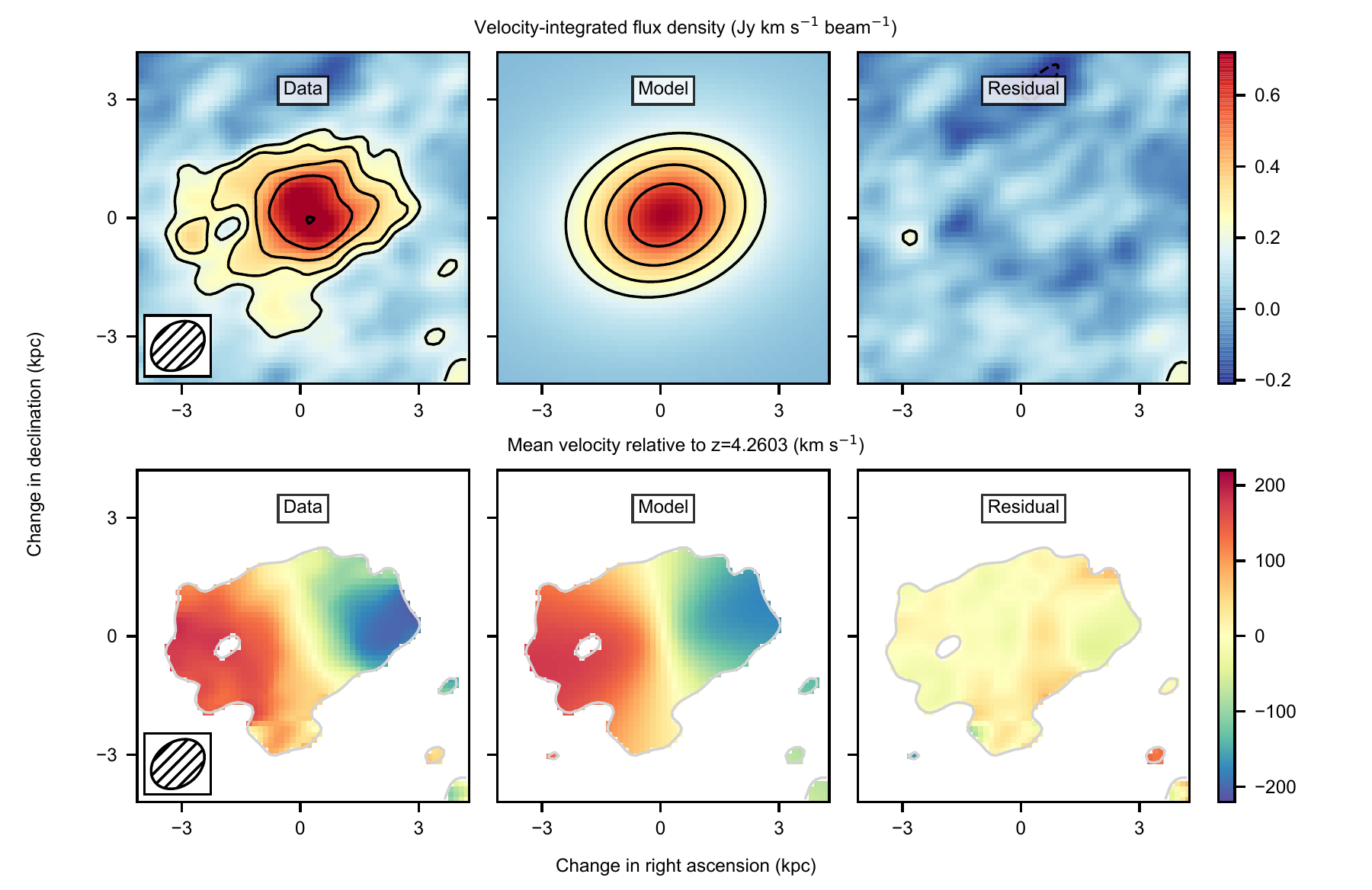}
\end{figure}
\noindent {\bf Fig.~2. Comparison between the data and the model for DLA0817g.} The top row shows the velocity-integrated \CII\ flux density for the data (left panel), the constant rotational velocity model (middle panel) and the residual after subtracting the model from the data (right panel). The outer contour is at 3$\sigma$, where $\sigma = 0.0656$~Jy~\kms is the standard deviation of the noise in the observations, with contours increasing in powers of $\sqrt{2}$. No negative contours at the same levels are observed in the image. The synthesized beam of the observations is shown in the bottom left corner of the leftmost plot. The bottom row shows the mean velocity of the \CII\ emission, for the data (left panel), the model (middle), and the residuals (right). Velocities are relative to the systemic velocity of the \CII\ emission, corresponding to $z =$~4.2603.

\newpage
\begin{figure}[!t]
\centering
\includegraphics[width=0.65\textwidth]{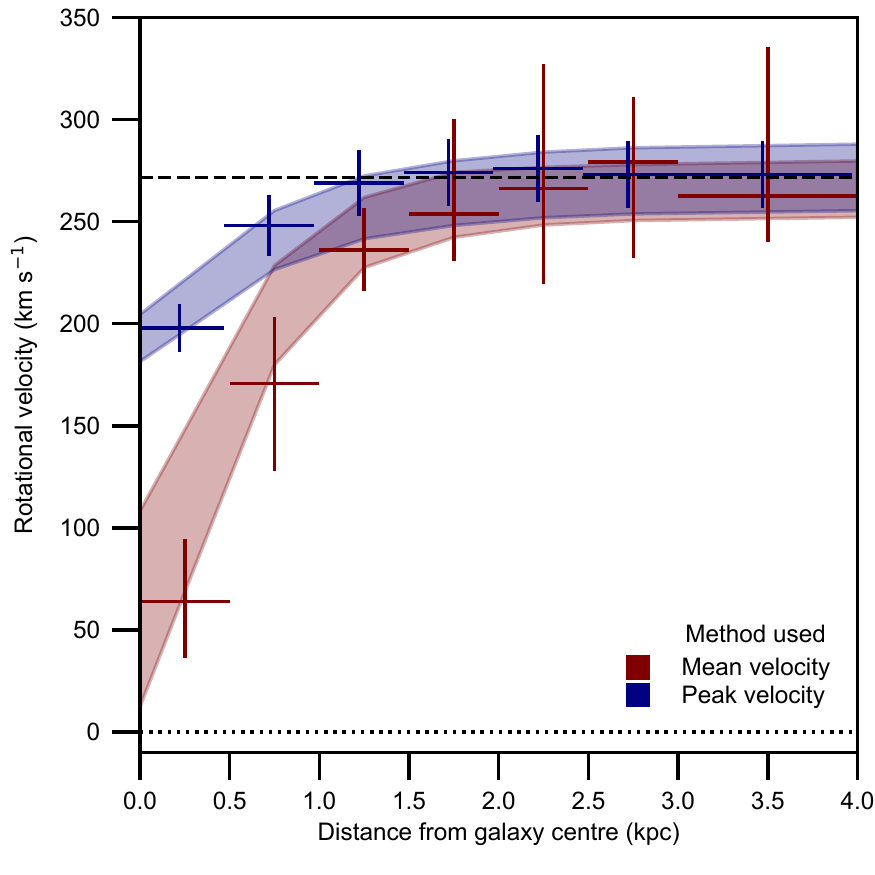}
\end{figure}
\noindent {\bf Fig. 3. Rotation curve for DLA0817g.} The rotation curve is derived from two different approaches, one approach (red curve) uses the mean velocity field to calculate the rotational velocity, and the other approach (blue curve) estimates the peak rotational velocity in the full data cube (Methods). The points show the observational data of the \CII\ emission line. The horizontal error bars indicate the size of the radial bins, whereas the vertical error bars indicate the 16 to 84 percentile range of the individual measurements within the bin. The shaded region marks the 16 to 84 percentile range when the same method is applied to the rotating disk model with constant velocity. The dashed line indicates a rotational velocity of 272~\kms, as estimated from the kinematic modelling. The decrease in the model and data curves below a distance of about 1.8~kpc is due to convolution with the ALMA beam, which has a size of approximately 1.3~kpc.

\newpage
\begin{figure}[!t]
\includegraphics[width=\textwidth]{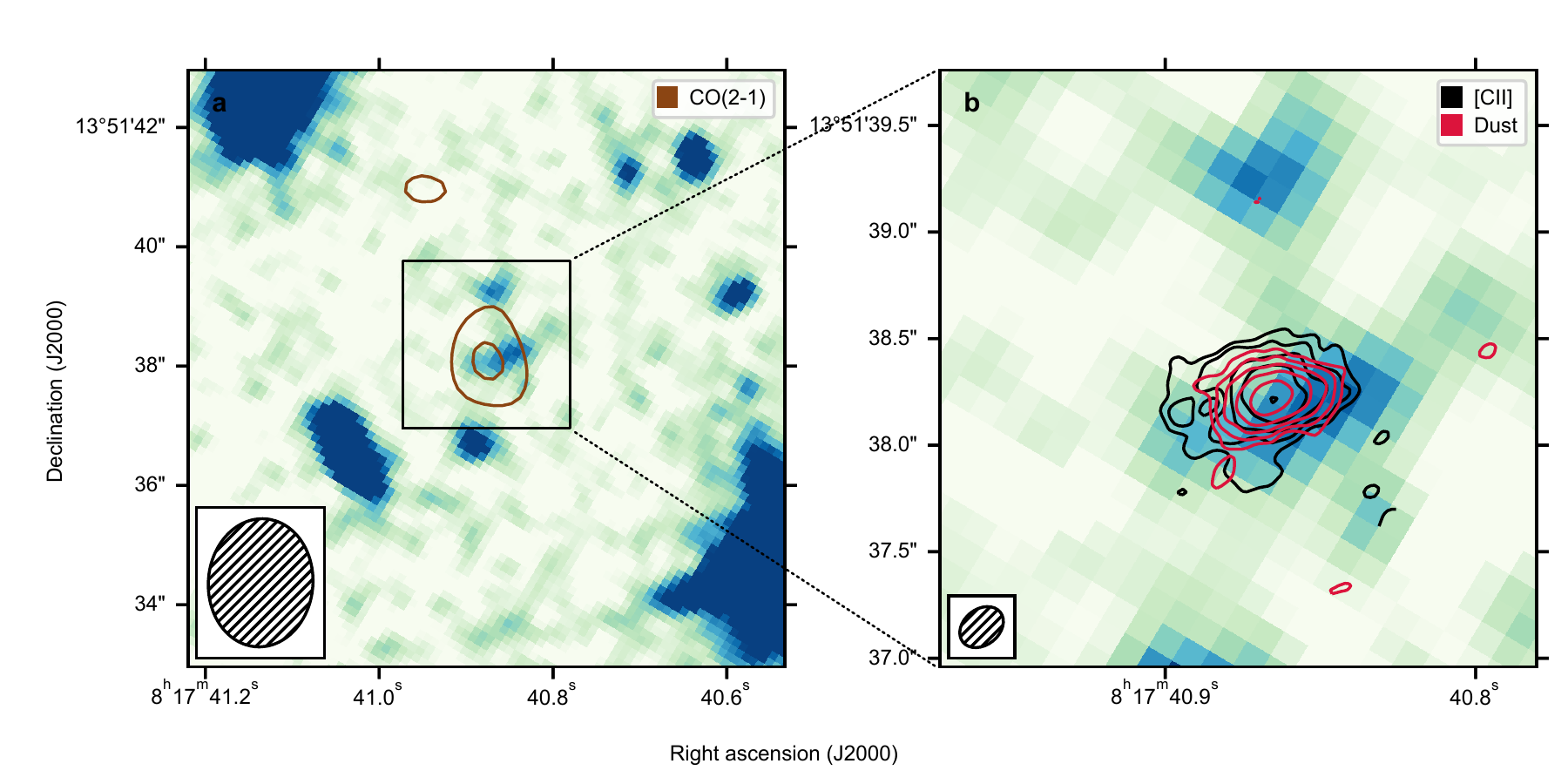}
\end{figure}
\noindent {\bf Fig. 4. HST imaging of DLA0817g with CO, \CII\ and dust continuum contours.} The F160W filter on HST's Wide Field Camera 3 was used to probe the rest-frame near-UV emission of the galaxy. {\bf a,} The field surrounding DLA0817g. The bright source in the bottom right corner is the quasar whose sightline contains the $z =$ 4.26 absorber. Overlaid on this figure in contours is the CO(2$-$1) emission obtained with the JVLA. Contours start at 3$\sigma$, where $\sigma = 36 \upmu$Jy beam$^{-1}$ is the stanard deviation of the noise in the observation, and increase in powers of $\sqrt{2}$. No negative contours are seen in this image at similar levels. The inset shows the synthesized beam of the JVLA observations. {\bf b,} An enlargement of the near-UV emission from DLA0817g. Overlaid on the figure are the contours of \CII\ (black) and dust continuum (red) emission obtained with ALMA. The synthesized beam of the ALMA observations is shown in the inset. Contours start at 3$\sigma$ and increase in powers of $\sqrt{2}$, where $\sigma = 0.0656$~Jy~\kms\ and 27 $\upmu$Jy beam$^{-1}$ for the \CII\ and dust continuum observations, respectively.

\newpage
\begin{methods}

\subsection{Cosmology.}
Throughout this paper we use a flat $\Lambda$ Cold Dark Matter cosmology, defined by the parameters, $\Omega_{\Lambda}=0.7$, $\Omega_{\rm M}=0.3$ and $H_0 = 70$~\kms~Mpc$^{-1}$. Adopting this concordance cosmology aids comparisons with previous results in the literature.

\subsection{ALMA observations and data reduction.}
ALMA observed the galaxy associated with the absorber toward QSO~J081740.52+135134.5, DLA0817g, in two executions on 2018~January~6 and 2018~January~21 (all dates are given in Universal Time) for a total on-source time of 100~minutes (programme ID 2017.1.01052.S, Principal Investigator (PI): Neeleman). One of the four 1.875-GHz ALMA receiver bands was centred at 361.312~GHz, the redshifted frequency of the [C\,\textsc{ii}]~158-$\upmu$m (\CII) emission line, with 240~channels, that is, a resolution of 7.8125~MHz per channel. The remaining three 1.875-GHz bands were set up for continuum observations at neighbouring frequencies. The blazar~J0854+2006 was used as a flux and bandpass calibrator, and the blazar~J0750+1231, as the phase calibrator. The nominal antenna configuration was C43-5, with a maximum baseline length of approximately 2.5~km. A total of 43 and 44 antennas were used for the two executions, respectively. 

The raw data were calibrated using the ALMA calibration pipeline, which is part of the Common Astronomy Software Application (CASA) package\cite{Mcmullin2007}. The calibrated visibility data set was then re-examined within CASA, where minor additional flagging was performed. From the line-free channels, we generated two continuum images by using the task \emph{tclean}, one using natural weighting, and the other using Briggs weighting with a robust parameter of 0.5. The resultant synthesized beams are 0.23$'' \times$ 0.17$''$ and 0.16$'' \times$ 0.12$''$ with root-mean-square (r.m.s.) sensitivities of 23~$\upmu$Jy beam$^{-1}$ and 27~$\upmu$Jy beam$^{-1}$, respectively. To generate the \CII\ emission line cube, we subtracted the continuum emission using the task \emph{uvsub} and then subtracted any remaining residual continuum emission with \emph{uvcontsub}. As with the continuum image, we made two different image cubes using \emph{tclean} for the two different weighting schemes resulting in r.m.s. sensitivities of 0.35~mJy~beam$^{-1}$ per 25~\kms\ for natural weighting and 0.41~mJy~beam$^{-1}$ per 25~\kms\ for Briggs weighting with a robust parameter of 0.5. In this paper, we opted to use only the natural weighting scheme, as the better sensitivity provided slightly better constraints than the higher resolution of the Briggs weighting scheme. However, none of the results presented here are affected by the choice of this weighting scheme.

\subsection{JVLA observations and data reduction.}
To search for redshifted CO(2$-$1) emission from DLA0817g, we observed the galaxy using the JVLA in four separate observing runs between 2017~March~03 and 2017~April~10, for a total on-source integration time of 9 h (programme ID 17-279, PI: Neeleman). One of the two JVLA intermediate frequency bands was centred on the redshifted CO(2$-$1) line at 43.828~GHz, with the three central sub-bands having a resolution of 250~kHz. The remaining sub-bands were set up at a coarser resolution of 1~MHz. The most compact configuration, D, was used, resulting in a synthesized beam of 2.1$'' \times$ 1.8$''$. We performed the calibration of the raw data using standard routines within CASA. The calibrated individual runs were combined using the task \emph{concat} and then a continuum image was made from the line-free channels, using the task \emph{tclean}. No continuum emission was detected within the primary beam of the observations, which covers both the galaxy, DLA0817g, at $z=4.2601$, and the quasar, SDSS~J081740.52+135134.5, at $z=4.398$. Two emission line cubes were created of the redshifted CO(2$-$1) line from DLA0817g with \emph{tclean} and natural weighting, one with a channel width of $50$~\kms, and the other with a width of $550$~\kms, with RMS sensitivities of 120~$\upmu$Jy beam$^{-1}$  and 36~$\upmu$Jy beam$^{-1}$, respectively. The latter cube was used to measure the CO(2$-$1) line flux density, and create the contours in Fig.~4.

\subsection{HST observations, data reduction and analysis.}
HST observations of DLA0817g (programme ID 15410, PI: Neeleman) were obtained on 2018~May~07 and consist of one orbit of four 653-s exposures with the Wide Field Camera~3, utilizing the F160W filter. This filter covers the rest-frame near-UV stellar light from the galaxy. We created the image mosaic using AstroDrizzle and aligned the astrometry to the GAIA DR2 astrometry\cite{Gaia2018} using TweakReg, resulting in an absolute astrometric uncertainty of approximately 0.006$''$. The effective spatial resolution of the HST observations is $0.31''$, as determined from a Gaussian fit to the point spread function of the quasar.

We detect rest-frame near-UV emission from DLA0817g at 7$\sigma$ significance within the isophotal area of the galaxy (22 pixels). To determine the photometry and basic shape of the stellar light from DLA0817g, we used Source Extractor (v.2.5.0\cite{Bertin1996}). The galaxy is extended with an ellipticity of 0.6 and a position angle of 114$^\circ \pm$ 8$^\circ$, fully consistent with the position angle determined from the kinematic analysis of the \CII\ line emission (see kinematic analysis section below). We measure the total flux using Source Extractor's {\tt flux\_auto}, which provides the flux within an elliptical aperture with the Kron radius\cite{Kron1980}. This yields a total AB magnitude of $M_{\rm AB} =$ 25.1 $\pm$ 0.2 and a Kron radius of 0.9$''$.

\subsection{Far-Infrared Continuum, \CII\ Line, and CO(2$-$1) Line Luminosities.}
The rest-frame 160 $\upmu$m continuum flux density of DLA0817g is detected and resolved by the ALMA observations with a flux density within a 1.5$''$ radius centered on the galaxy of 1.28 $\pm$ 0.15~mJy. The size of the emission is (0.41 $\pm$ 0.05$''$ $\times$ 0.24 $\pm$ 0.03$''$), as determined from a Gaussian fit to the data. These measurements are consistent with the values obtained from the lower-resolution measurement\cite{Neeleman2017}, indicating that no emission is resolved out by the higher-resolution ALMA observations. From this single far-infrared measurement, we can estimate the total far-infrared luminosity ($L_{\rm TIR}$, defined as the integrated luminosity between 8 and 1000~$\upmu$m), assuming that the emission has a modified blackbody spectrum, and with the caveat that neither the dust temperature ($T_{\rm d}$) nor the dust emissivity power law spectral index ($\beta$) are constrained by this single measurement. Assuming fiducial values of $T_{\rm d} =$~35~K and $\beta =$~1.6, with possible ranges of 25~K $< T_{\rm d} <$ 45~K, and 1.2~$< \beta <$ 2.0, gives a total far-infrared luminosity estimate of $L_{\rm TIR}$ = 1.2$^{+2.3}_{-0.7}~\times$ 10$^{12}~L_\odot$, consistent with previous estimates\cite{Neeleman2017}. All observational measurements are listed in Extended Data Table 1.

To estimate the total flux from the \CII\ emission, we measure the flux density within a radius of 1.5$''$ centered on DLA0817g for each channel. The resultant spectrum is shown in Extended Data Fig.~1. To estimate the velocity-integrated line flux density, we integrated the spectrum over the channels showing \CII\ emission (marked by the horizontal bar in Extended Data Fig.~1). The total velocity-integrated \CII\ flux density from DLA0817g is 5.8~$\pm$~0.4~Jy~\kms\ resulting in a \CII\ luminosity of $L_{\rm [C\,{\scriptscriptstyle II}]}$~=~(3.26~$\pm$~0.22)~$\times$~10$^{9} L_\odot$. This value is consistent with that obtained from the lower-resolution data\cite{Neeleman2017}, indicating that the \CII\ emission is not resolved out in the higher resolution observations. To estimate the extent of the emission we fit a Gaussian profile to the velocity-integrated map of the \CII\ emission, resulting in a size of (0.63 $\pm$ 0.07$''$ $\times$ 0.43 $\pm$ 0.05$''$).

An emission line is detected in the JVLA data cube at 4.7$\sigma$ significance, with a velocity-integrated flux density of 94~$\pm$~20~mJy~\kms. Both the spatial location and the velocity of this spectral feature are in excellent agreement with the position and velocity of the \CII\ emission line from DLA0817g, strengthening the identification of the feature as the CO(2$-$1) line from the galaxy. Taking into account that the observations are taken against the cosmic microwave background (CMB) results in a correction factor of 1/(1-$B_\upnu[T_{\rm CMB}] / B_\upnu[T_{\rm exc}])$, where $T_{\rm CMB}$ is the temperature of the CMB at the redshift of DLA0817g, $T_{\rm exc}$ is the excitation temperature of the CO(2$-$1) line, and $B_\upnu[T]$ is the black body intensity at temperature $T$, and the frequency of the CO(2$-$1) line\cite{DaCunha2013}. As the excitation temperature is not known, we take a range of values between 25~K and 45~K, resulting in a correction factor of 1.5$\pm$~0.3 and an intrinsic velocity-integrated CO(2$-$1) flux density of 0.14~$\pm$~0.04~Jy~\kms. Converting the velocity-integrated flux density into a line luminosity using standard equations\cite{Carilli2013} gives a line luminosity of $L'_{\rm CO(2-1)}$~=~(2.4~$\pm$~0.7)~$\times$~10$^{10}$~K~\kms~pc$^2$ or $L_{\rm CO(2-1)}$ = (9.6~$\pm$~2.8)~$\times$~10$^6~L_\odot$.

\subsection{Star Formation Rate Estimates.}
We can convert both the far-infrared dust continuum measurement and \CII\ luminosity into a star formation rate (SFR) estimate. Dust continuum emission is expected to be a good tracer of the SFR as the far-infrared emission arises from dust that has been heated by stellar ultraviolet photons. To estimate the SFR from the far-infrared continuum measurement, we use the observationally-determined calibration between SFR and 160~$\upmu$m continuum\cite{Calzetti2010}, with the caveat that a fraction of the emission might arise from a population of older stars not associated with the current SFR\cite{Lonsdale1987}. However, this fraction is small for galaxies with comparable dust continuum\cite{Calzetti2010}. The measured star formation rate from the 160~$\upmu$m continuum is SFR$_{\rm 160\upmu m}$ = 118 $\pm$ 14~$M_\odot$~yr$^{-1}$. This value is in rough agreement with the SFR as determined from converting the total infrared luminosity (TIR) into a SFR using the scaling relationship given for local galaxies\cite{Kennicutt2012}: SFR$_{\rm TIR}$ = 177$^{+344}_{-103}~M_\odot$~yr$^{-1}$. Finally, observations at both low and high redshifts have shown a relationship between SFR and \CII\ luminosity\cite{Delooze2014,Herrera-Camus2015}. Using this relationship gives a SFR of SFR$_{\rm [C\,{\scriptscriptstyle II}]}$ = 420 $\pm$ 260~M$_\odot$~yr$^{-1}$. This is higher than the previous estimates, but in agreement with published results that show that DLA0817g sits slightly above this locally-derived relationship\cite{Neeleman2017}.

Using the rest-frame far-ultraviolet flux measurement, we can also provide an estimate of the SFR which is not obscured by dust. By applying the scaling relationship for local galaxies and assuming a Kroupa IMF\cite{Kennicutt2012}, the 1.6$\upmu$m measurement of DLA0817g corresponds to SFR$_{\rm 1.6 {\upmu}m}$~=~16~$\pm$~3~$M_\odot$~yr$^{-1}$. The large discrepancy between the dust-obscured SFR estimate and the far-ultraviolet SFR estimate indicates that DLA0817g has a significant amount of dust, which obscures the ultraviolet radiation.

\subsection{Molecular Gas Mass Estimates.}
Several estimators have been used in the literature to estimate the molecular gas mass of high-redshift galaxies. The CO(2$-$1) luminosity can be converted into a molecular gas mass estimate by assuming two quantities: the luminosity ratio between the CO(2$-$1) and CO(1$-$0) emission lines, r$_{\,2,1}$, and the conversion factor between the CO(1$-$0) luminosity and the molecular gas mass, $\alpha_{\rm CO}$. If the CO emission lines were thermalized, then r$_{\,2,1}$ would be equal to 1. However, this is unlikely to be the case, because star-forming galaxies at lower redshifts are slightly sub-thermally populated\cite{Dessauges-Zavadsky2015}. We therefore assume r$_{\,2,1} =$~0.81, as determined from these galaxies. The other parameter, $\alpha_{\rm CO}$ is less well-determined, and can vary between values of $\approx$1~$M_\odot$~(~K~\kms~pc$^2$)$^{-1}$ for quasars and starburst galaxies, to $\approx$3.0-4.3~$M_\odot$~(~K~\kms~pc$^2$)$^{-1}$ for the Milky Way and other disk galaxies, to $\gg$10~$M_\odot$~(~K~\kms~pc$^2$)$^{-1}$ for low-metallicity galaxies\cite{Bolatto2013}. As DLA0817g has physical properties similar to those of regular star-forming galaxies, and not the extreme conditions present in rapidly star-forming quasars and starburst galaxies nor the low SFR and metallicity of dwarf galaxies, we assume a conservative $\alpha_{\rm CO}$ of 3.0~M$_\odot$~(~K~\kms~pc$^2$)$^{-1}$ as was measured for color-selected galaxies at high redshift\cite{Daddi2010}. The resultant molecular mass estimate is $M_{\rm mol, CO} =$~ (8.8~$\pm$~2.6)~$\times$~10$^{10}$~$\times$~(0.81 / r$_{\,2,1}$)~$\times$~($\alpha_{\rm CO}$ / 3.0)~$M_\odot$.

The molecular gas mass of a galaxy may also be estimated via two other, more indirect, methods, using conversion factors from the far-infrared continuum luminosity\cite{Scoville2014} or the \CII\ line luminosity\cite{Zanella2018} to the molecular gas mass. The mass estimates from these two methods are $M_{\rm mol, FIR} =$~(5.7 $\pm$ 0.7) $\times$ 10$^{10}$ $\times$ (6.7 $\times$ 10$^{19}$ / $\alpha_{850\upmu {\rm m}})~M_\odot$ and $M_{\rm mol, [C\,{\scriptscriptstyle II}]} =$~(9.8 $\pm$ 0.6) $\times$ 10$^{10} \times (\alpha_{\rm [C\,{\scriptscriptstyle II}]}$/30)~$M_\odot$, respectively. The results from both methods are in agreement with the molecular gas mass estimate based on the CO(2$-$1) emission line. 

For all molecular mass estimates we only report the observational uncertainties. The uncertainties due to the conversion factors are considered in the measurement by reporting the molecular mass as a function of the conversion factor. For both the far-infrared continuum-based and the \CII\ line luminosity-based molecular gas mass estimates, these uncertainties have not been studied in detail as a function of galaxy properties, but are presumed to be larger than the observational uncertainties; e.g., for the \CII\ conversion factor, $\alpha_{\rm [C\,{\scriptscriptstyle II}]}$, the scatter is at least a factor of 2\cite{Zanella2018}, while a similar scatter of $\sim$50~\% is observed in the far-infrared continuum conversion factor, $\alpha_{850\upmu {\rm m}}$\cite{Scoville2014}. The conversion factor from the CO(2$-$1) emission line to molecular mass is a function of the galaxy's star formation and metallicity\cite{Bolatto2013}, and is better calibrated than the other measurements. We therefore report the molecular gas mass estimate based on the CO(2$-$1) emission line.

\subsection{Kinematic Analysis of the \CII\ Emission Line.}
To model the dynamics of the \CII\ emission line, we have fitted the observed data cube using a custom Python program that generates a model cube from a user-defined emission model. The program convolves the model cube with the observed beam and then minimizes the residuals between the convolved model and the observed data cube using a Markov Chain Monte Carlo (MCMC) approach\cite{Banados2019,Neeleman2019b,Venemans2019}. This program also yields estimates of the uncertainty on each of the parameters of the model.

For DLA0817g, we assume a \CII\ line emission model in which the \CII\ emission arises from a thin disk in which the emission exponentially declines: $I(R) = I_0 e^{-R/R_{\rm d}}$.  The velocity curve of the \CII\ emission is assumed to either be constant as a function of radius, $v(R) = v_{\rm rot}$, or increase with radius via an arctangent profile, $v(R) =  2v_{\rm rot}/\uppi \arctan(R/R_{\rm v})$. In both models, we assume a constant velocity dispersion across the disk, $\sigma(R) = \sigma_{\rm v}$. Together with the three spatial coordinates ($x_0, y_0, z_0$), and the inclination ($i$) and position angle ($\alpha$), these 9 or 10 parameters uniquely determine the \CII\ line emission in the model cube.

The results of the MCMC analysis are given in Extended Data Table~2. No significant differences are found between the two different models, and therefore we opt for the model with the least free parameters, the constant rotational velocity model. The top-left and top-middle panels of Fig.~2 show the velocity-integrated \CII\ flux density of DLA0817g for the observed data and the model with constant rotational velocity. The top-right panel shows the residual between the model and the data, which lacks any $\geq$ 3$\sigma$ significance features. The bottom panels show the velocity field of the model and observations. For completeness, we show the channels maps of DLA0817g and the best fit model (Extended Data Figs.~2 and 3) and the position-velocity diagrams along the major and minor axis (Extended Data Fig.~4). Little excess emission (with $\geq$ 3$\sigma$ significance) is seen in the channel maps of the residual, suggesting that the simple exponential thin-disk, constant rotational velocity model is an accurate representation of the bulk of the \CII\ emission.

In the above models, we assume the \CII\ emission arises from an infinitely thin disk. This is a simplified assumption, as disks at high redshifts are expected to be turbulent and thus thicker than local analogues\cite{Bird2013}. The high velocity dispersion in DLA0817g, compared to measurements at $z \approx 2$ from H$\upalpha$ emission\cite{Simons2017,Forster-Schreiber2009} is a further indication that the gas disk is likely turbulent. To explore how the thin-disk approximation affects the kinematic analysis, we repeated the modeling for models where the \CII\ distribution is distributed in a thick disk with an exponential scale height whose scale radius is varied from 0.15 to 1.0 times the scale length in the radial direction. These models have systematically lower inclinations by an average of 7$^\circ$, and thus higher rotational velocities by $\sim$50~\kms. Furthermore, the velocity dispersion estimates are lower by 10 \kms, as some of the dispersion arises from line-of-sight motion in the thick disk. Combining both results would increase $v_{\rm rot} / \sigma_v$ by 30~\%. As no information is available about the thickness of the disk, we take a conservative approach and report values based on the thin-disk approximation, where the uncertainties include the spread due to the thickness of the disk. As a consistency check, we also fitted DLA0817g using the 'tilted-ring' fitting program $^{\rm 3D}$Barolo\cite{DiTeodoro2015}, which yields a rotational velocity of  279~\kms and a velocity dispersion of 72~\kms, consistent with our values. 

\subsection{Dynamical Mass Estimate.}
To estimate the dynamical mass, we assume that the gas is rotationally supported.  With this assumption, the dynamical mass ($M_{\rm dyn}$) of a system (in $M_\odot$) within a radius $R$ (in kpc) is given by: $M_{\rm dyn}(R)$ = 2.32 $\times$ 10$^{5} v_{\rm rot}(R)^2 R$, where $ v_{\rm rot}(R)$ is the circular velocity (in \kms) of the galaxy at the radius $R$\cite{Wang2013}. We note that in deriving this equation, the mass distribution is implicitly assumed to be spherical. For an exponential thin-disk mass distribution, this underestimates the mass by up to 30\%, depending on the radius at which the rotational velocity is measured\cite{Walter1997}.

Our kinematic analysis provides an estimate for the --constant-- rotational velocity, $v_{\rm rot} = 272^{+52}_{-13}$~\kms. We can compare this rotational velocity estimate to estimates of the rotational velocity from other estimators used in lower resolution data, such as the full width at half maximum (FWHM) of the integrated \CII\ emission line. By fitting a double Gaussian profile to the integrated \CII\ spectrum (Extended Data Fig.~1), we measure an FWHM$_{\rm [C\,{\scriptscriptstyle II}]}$ of 400~$\pm$~40~\kms. If we assume the emission arises from ordered rotation and can be described by a sharp double Gaussian profile, then $v_{\rm rot} $ = 0.5 $\times$ FWHM$_{\rm [C\,{\scriptscriptstyle II}]} / \sin(i)$\cite{Ho2007,Wang2013}. With the inclination estimate of $i$ = 42$^\circ$ $\pm$ 3$^\circ$, we get a rotational velocity of 300 $\pm$ 50~\kms. This estimate is consistent with the rotational velocity estimate derived from our kinematic modeling, but we note that estimating the rotational velocity from the integrated line spectrum requires both an accurate inclination and a resolved spectral line profile.

To measure a dynamical mass for a galaxy, we must define a radius at which to measure this mass. Most high-redshift galaxy observations barely resolve the emission, and the radius chosen is either the maximum extent of the emission\cite{Wang2013}, or the major semi-axis of the two-dimensional Gaussian fit to the \CII\ emission\cite{Decarli2018}. In our resolved observations, we will assume the extent of the galaxy to be three times the exponential scale length of the modeled emission. This extent ($R$ = 4.2~kpc~= 0.6$''$) corresponds to a region that emits 80~\% of the total \CII\ flux density of the galaxy. It is also similar to the maximum extent of the \CII\ line emission that yields a reliable rotational velocity measurement (Fig. 2). Within this region, we measure a dynamical mass of $M_{\rm dyn}$ =  (7.2 $\pm$ 2.3) $\times$ 10$^{10} M_\odot$, where the uncertainty includes, in quadrature, a 30\% uncertainty from the assumption of a spherical mass distribution.

\subsection{Mean Velocity Field, Velocity Curves and Light Profiles.}
To estimate the mean velocity field for DLA0817g (Fig.~1), we fit a Gaussian profile to the spectrum of each spatial position where the velocity-integrated flux is detected at  $>$3$\sigma$. The mean of the Gaussian fit is the estimated mean velocity at that position. This method is more robust against flux outliers than the standard first moment map\cite{DeBlok2008}. Using this velocity field and the de-projected distance of the pixel --determined from the inclination, galaxy center, and position angle from the kinematic modeling-- yields an estimate of the rotation velocity at each pixel. Here we assume that all of the velocity is constrained within the plane of the galaxy. The resultant median and 1$\sigma$ spread in data points per radial bin are shown in Fig.~3 in red, and labeled as the mean velocity method. For the second method, we take the full data cube and de-project the cube into the plane of the galaxy, again assuming that all of the velocity is constrained within the plane of the galaxy. We then take the intensity of each 3D-pixel and its associated rotational velocity as the probability that the rotational velocity has that value at that position. By averaging this over all the 3D pixels within a radial bin, we get a probability distribution function  (PDF) of the rotational velocity per radial bin. The rotational velocity per bin is taken to be the peak of a spline fit to the PDF and is shown by the blue points in Fig.~3, labeled as the peak velocity method. In both cases, uncertainties are convolved with the uncertainties in both the inclination and position angle. The peak of a distribution is less affected by asymmetric profiles, such as those arising from beam-smearing, explaining why the peak velocity method yields better constraints closer to the center of the galaxy. 

To provide a validity check for our measurement of the velocity dispersion, we create a velocity dispersion radial profile. This profile is generated from the measurements of the width (i.e., standard deviation) of the fitted Gaussian profiles, and the de-projected distance at each pixel. The resultant resultant median and 1$\sigma$ spread in data points per radial bin are shown in Extended Data Fig.~5. The horizontal dashed line indicates the value for the velocity dispersion as determined from the kinematic modeling. This figure shows that away from the kinematic center, where beam-smearing is less severe, the observed velocity dispersion is in agreement with the velocity dispersion determined from the kinematic modeling. 

The light profiles for the dust continuum, \CII\, and near-UV emission are shown in Extended Data Fig.~6. The dust and near-UV continuum emission light profiles are directly measured from the continuum maps of the ALMA and HST data, respectively. For the \CII\ emission we create an integrated flux density map over the channels highlighted in Extended Data Fig.~1 (see also Fig.~2). We do not attempt a CO(2$-$1) light profile, as the emission is not resolved. Before generating the light profiles, the \CII\ and dust continuum are convolved with a Gaussian kernel to account for the slightly worse resolution of the near-UV data. All of the light profiles are plotted against the de-projected radius, assuming the same inclination, position angle and kinematic center. The points are radially binned and the measurements include systematic uncertainties from taking the same position and orientation between the different emission profiles. These measurements show that the dust is more compact than the \CII\ emission, whereas the near-UV is consistent within the uncertainties with the \CII\ emission.

\subsection{Toomre-$Q$ Parameter.}
The Toomre-$Q$ parameter is a quantity that defines how stable a system is against gravitational perturbations\cite{Toomre1964}. For a collisionless gas, this parameter is given by the equation: $Q = \sigma_v \kappa / \pi G \Sigma_{\rm gas}$\cite{Goldreich1965},  with $Q <$ 1 corresponding to gas that is unstable. In this equation, $\kappa$ is the epicyclic frequency, which, for a constant-velocity thin disk, is given by $\kappa = \sqrt{2} v_{\rm rot} / R$, where $\Sigma_{\rm gas}$ is the gas surface density. If we assume that the \CII\ emission line traces the gas surface density, then --together with the total molecular gas mass derived from the CO(2$-$1) line-- we can estimate the gas surface density. The spatially-resolved, beam-convolved Toomre-$Q$ parameter distribution for DLA0817g is shown in panel~A of Extended Data Fig.~7. 

Although the \CII\ emission line is resolved, the limited resolution (i.e., beam-smearing) affects the measurement of $Q$. To explore the effect of resolution on the measurement of $Q$, we can estimate an azimuthally-averaged $Q$ directly from the surface density profile found in the kinematic modeling, which takes into account the limited resolution of the observations. The resultant radial profile for $Q$ is shown by the black line in panel B of Extended Data Fig.~7. This resolution-independent measurement shows that at all radii beam-smearing causes $Q$ to be systematically lower, because the emission is spread out over a larger region. To estimate a net Toomre-$Q$ parameter for the whole galaxy, $\bar{Q}$, we average the radial profile for $Q$ over the extent of the emission (R = 3$R_{\rm d}$) according to
\begin{equation}
\bar{Q} = \frac{2\uppi\int_{0}^{3R_{\rm d}} Q(R) R dR}{2\uppi\int_{0}^{3R_{\rm d}} R dR} = \frac{4\sqrt{2}\sigma_vv_{\rm rot}R_{\rm d}}{9GM_{\rm tot}}(e^{3} - 1).
\end{equation}
Here, the last equality holds because the gas surface density profile, $\Sigma_{\rm gas}(R)$ is a simple exponential and the integral can be solved analytically. This yields $\bar{Q} = 0.96 \pm 0.30$ for DLA0817g. 

We note that we only calculate the Toomre-$Q$ parameter for the gas. The total $Q$ parameter should also include contributions from other galaxy constituents, in particular the stars. However,  the total $Q$ parameter is the inverse of the sum of the inverses of the individual $Q$ parameters. As such, given that the disk is already marginally unstable from the gas contributions alone, any additional component will only make the disk more unstable. Our estimate of the $Q$ parameter can therefore be taken as an upper limit. Another important caveat is that these measurements are averages over the size of the beam, and variations ---especially in the surface density--- on smaller scales could mean that certain regions within the disk are stable. On smaller scales, the gas surface density is also not likely to be traced very well by the \CII\ emission, especially in the higher density regions, where either the \CII\ emission is self-absorbed\cite{Riechers2013}, or most of the carbon is in the neutral state or locked up in CO\cite{Beuther2014}. However, these regions will be unstable and the measured $Q$ parameter can again be taken as an upper limit.  

\subsection{Data Availability.}
The data reported in this paper are available though the ALMA archive:\\ (http://almascience.eso.org/aq/) with project code: 2017.1.01052.S, the JVLA archive:\\ (https://science.nrao.edu/facilities/vla/archive/index) with project code: 17A-279, and the HST/ Mikulski Archive for Space Telescopes: (https://archive.stsci.edu/hst/) with project code: 15410.

\subsection{Code Availability.}
All of the code used to generate the kinematic modeling is available online at: (https://github.com/mneeleman/qubefit).\\

\end{methods}

\begin{addendum}
 \item This work would not have been possible without the insights of the late Arthur M. Wolfe. M.N. thanks F. Walter for discussions, and S. Simha for extracting effective radii from the HST and ALMA imaging. ALMA is a partnership of the European Southern Observatory (ESO; representing its member states), the National Science Foundation (NSF; United States) and the National Institutes of Natural Sciences (Japan), together with the National Research Council (Canada), the National Science Council and the Academia Sinica Institute of Astronomy and Astrophysics (Taiwan), and the Korean Astronomy and Space Science Institute (Republic of Korea), in cooperation with the Republic of Chile. The Joint ALMA Observatory is operated by ESO, Associated Universities Incorporated (AUI.) / National Radio Astronomy Observatory (NRAO) and the National Astronomical Observatory of Japan. NRAO is a facility of the NSF operated under cooperative agreement by AUI. M.N. acknowledges support from the European Research Council advanced grant 740246 (Cosmic\_Gas). N.K. acknowledges support from the Department of Science and Technology via a Swarnajayanti Fellowship (DST/SJF/PSA-01/2012-13), and from the Department of Atomic Energy under project 12-R\&D-TFR-5.02-0700. Support for program number 15410 was provided by the National Aeronautics and Space Administration (NASA) through a grant from the Space Telescope Science Institute, which is operated by the Association of Universities for Research in Astronomy Incorporated, under NASA contract NAS 5-26555.
 \item[Author Contributions] M.N. is the PI of the observing programmes. M.N.  and N.K. contributed to the analysis of the ALMA and JVLA data, M.R. reduced and analysed the HST data. All authors contributed to writing and editing of the manuscript.
 \item[Competing Interests] The authors declare no competing interests.
 \item[Correspondence] Correspondence and requests for materials should be addressed to M.N.~(email: neeleman@mpia.de).
\end{addendum}

\newpage
\begin{figure}[!t]
\centering
\includegraphics[width=0.7\textwidth]{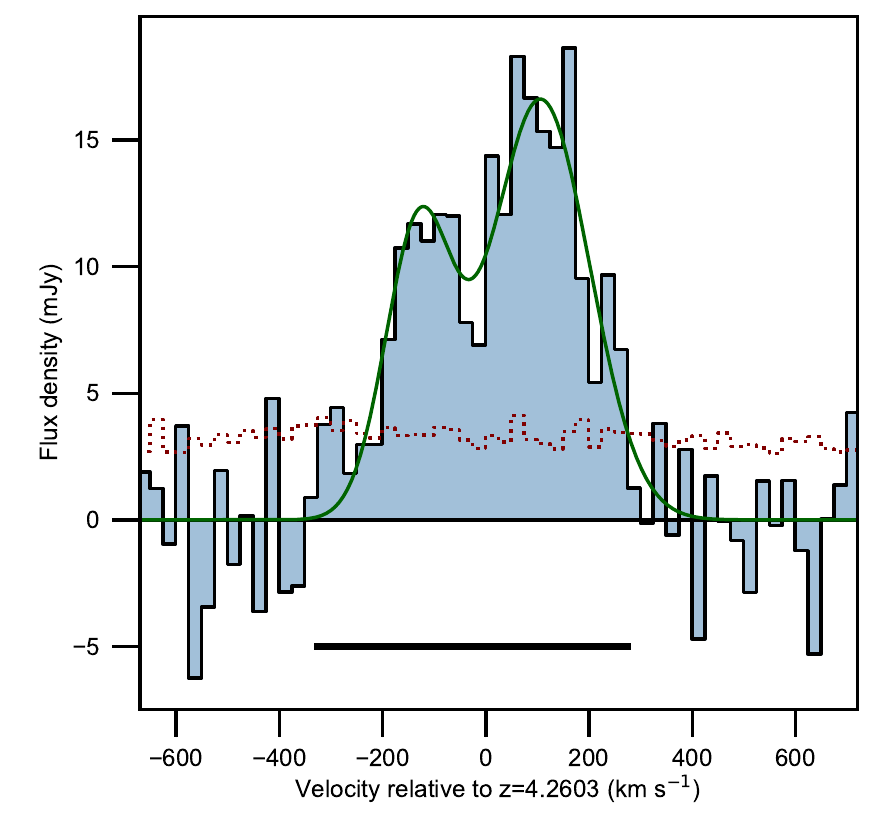}
\end{figure}
\noindent {\bf Extended Data Fig.~1. Spectrum of the \CII\ line from DLA0817g.} The velocity is relative to the systemic velocity of the \CII\ line. The velocity range used to estimate the velocity-integrated \CII\ flux density and the integrated \CII\ contours is marked by the solid black bar. The 1$\sigma$ uncertainty of the measurements is indicated by the dotted red lines, and has been estimated by bootstrapping flux density measurements at random positions within each channel chosen to be devoid of any line emission. A double Gaussian model fit to the data is shown in green. Both the peak flux density of 16.8~$\pm$~1.3~mJy and the velocity-integrated \CII\ line flux density of 5.8~$\pm$~0.4~Jy~\kms\ are consistent with values obtained from the lower resolution data, indicating that no emission is resolved out by the higher resolution observations.

\newpage
\begin{figure}[!t]
\centering
\includegraphics[width=0.8\textwidth]{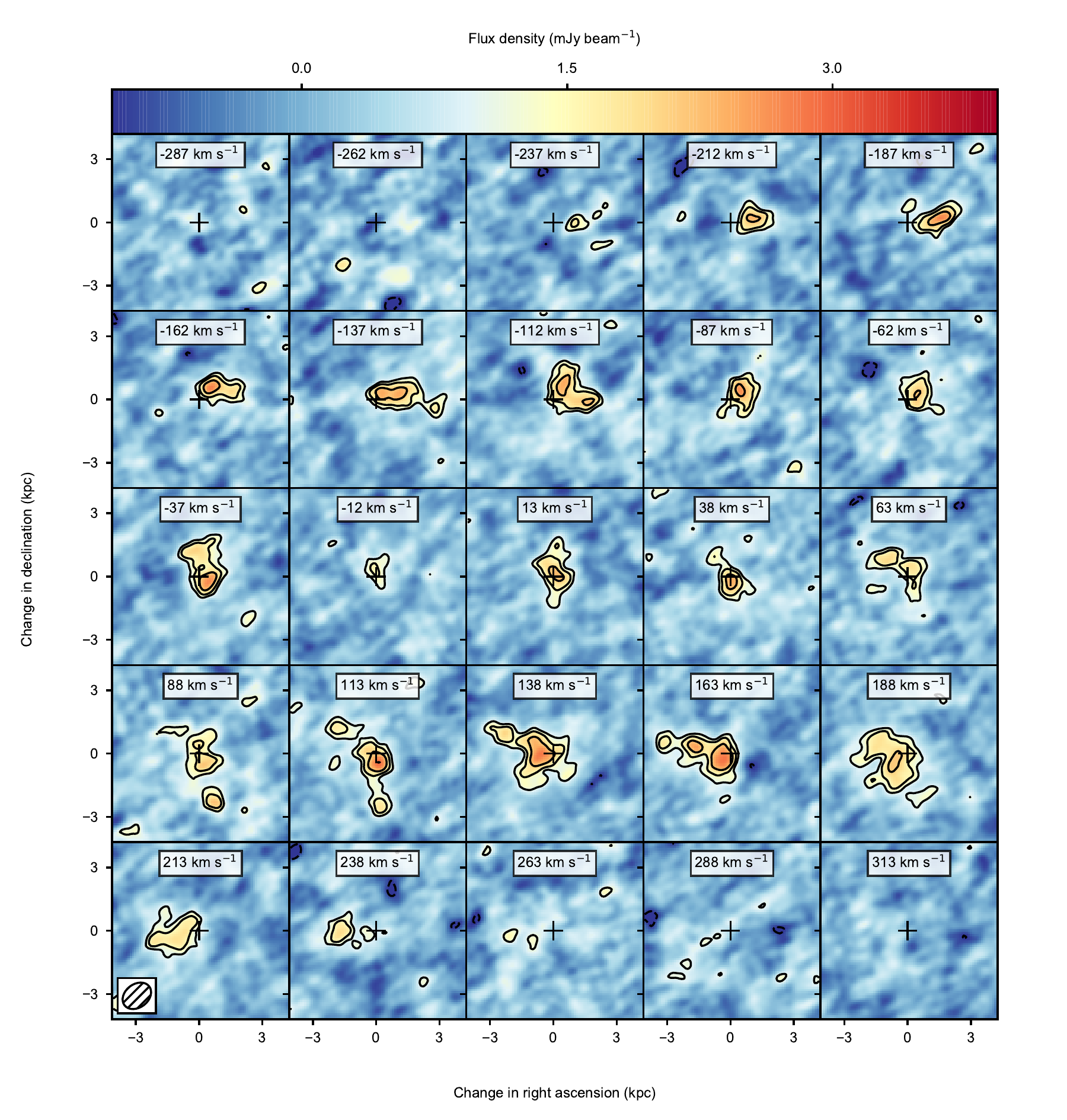}
\end{figure}
\noindent {\bf Extended Data Fig.~2. Channel maps of the \CII\ emission line from DLA0817g.} The plus symbol indicates the central position of the [C\,\textsc{ii}] emission derived from the kinematic analysis. This agrees within 1$\sigma$ with the position derived from fitting a 2D-Gaussian profile to both the velocity-integrated \CII\ emission and the far-infrared continuum emission, using the \emph{imfit} routine in CASA. The outer black contour is at $3\sigma$ significance, with subsequent contours increasing by powers of $\sqrt{2}$. Velocities are relative to the kinematically-derived \CII\ redshift, $z = 4.2603$. The synthesized beam is shown in the bottom left corner of the bottom left panel.

\newpage
\begin{figure}[!t]
\centering
\includegraphics[width=0.7\textwidth]{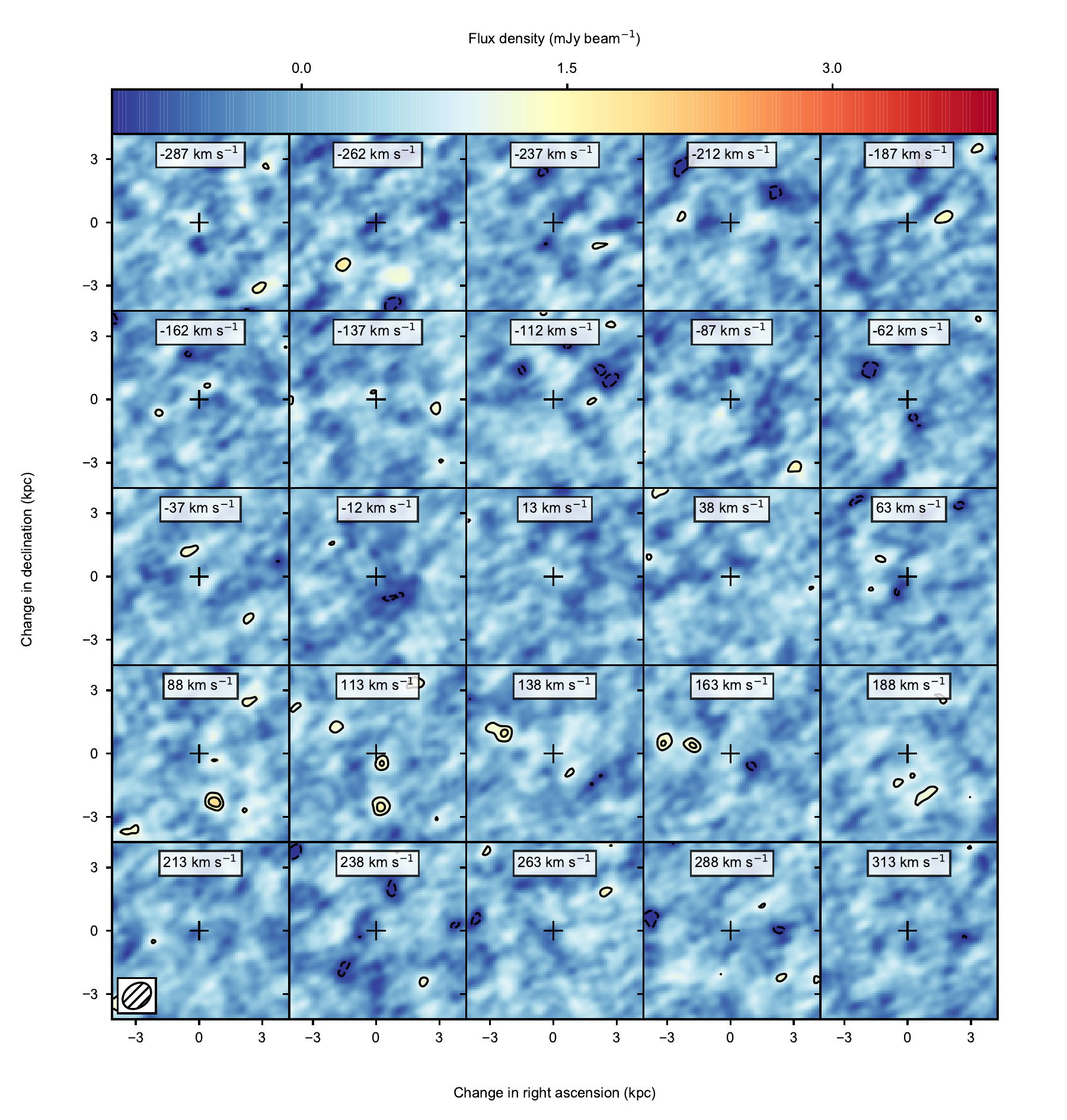}
\end{figure}
\noindent {\bf Extended Data Fig.~3. Channel maps of the residuals, after subtracting the model from the data, of the \CII\ emission from DLA0817g.} The color scaling and annotations are the same as Extended Data Fig.~2. Little excess emission (at $>3\sigma$ significance) is seen in the individual 25~\kms\ channels, indicating that the exponential thin disk model is a good approximation for the bulk of the \CII\ emission. Only two features are seen in the channel maps with $>$3$\sigma$ emission in two or more consecutive channels: 2.8~kpc south of the center at 88~\kms\ and 113~\kms, and 3~kpc east of the center at 113~\kms, 138~\kms, and 163 \kms. This emission arises from clumps that are not rotating with the bulk of the gas, possibly arising in outflows or satellite galaxies. 

\newpage
\begin{figure}[!t]
\includegraphics[width=\textwidth]{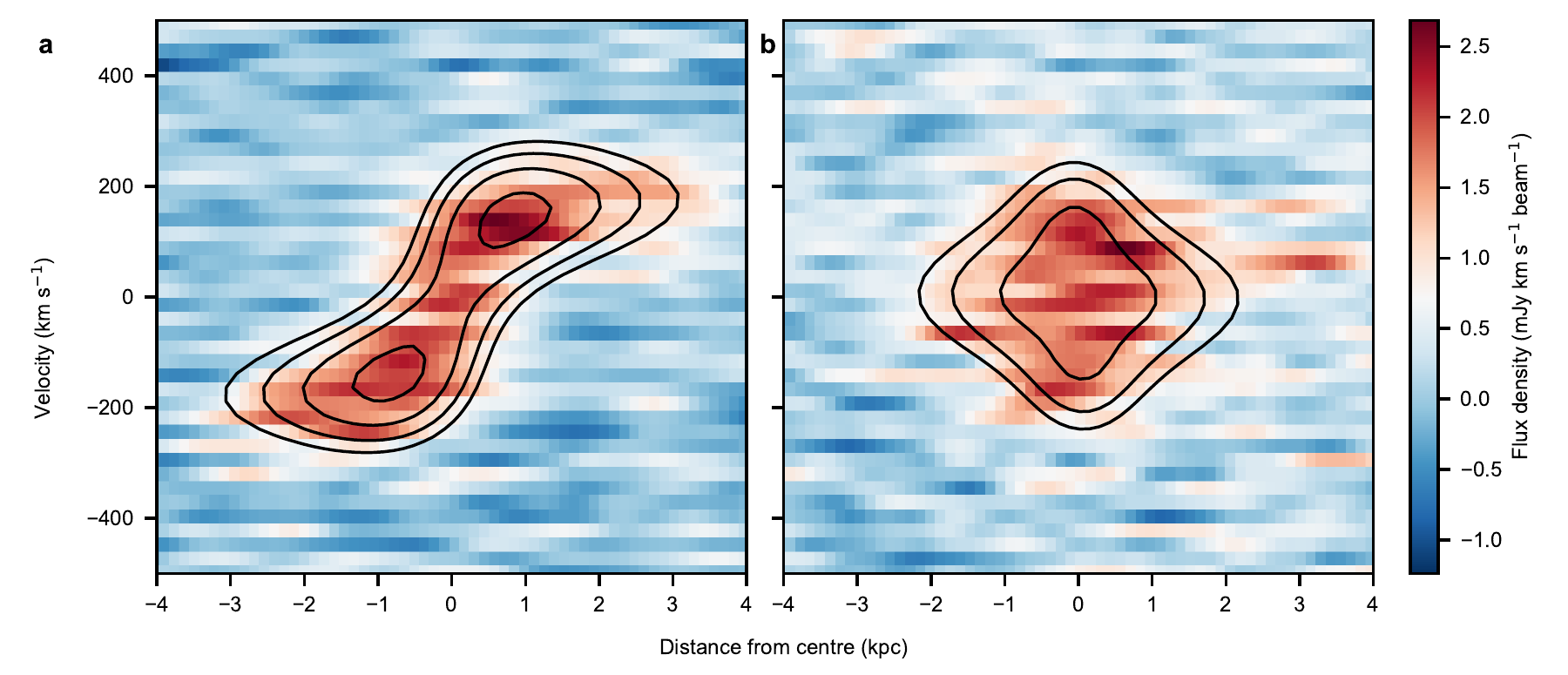}
\end{figure}
\noindent {\bf Extended Data Fig.~4. Position-velocity diagram for DLA0817g.} Panel A shows the position-velocity diagram along the major axis of DLA0817g, and panel B shows the position-velocity diagram along the minor axis. The contours in both panels are the position-velocity diagrams derived from the constant velocity model. Contours start at 2$\sigma$ and increase in powers of $\sqrt{2}$.

\newpage
\begin{figure}[!t]
\centering
\includegraphics[width=0.7\textwidth]{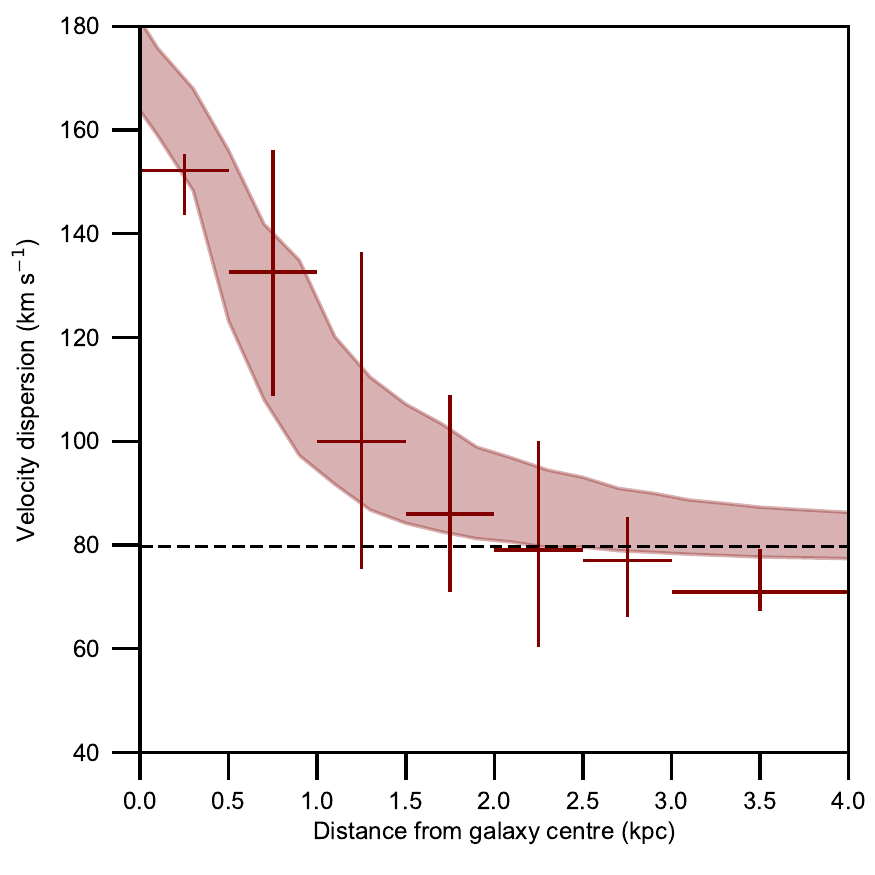}
\end{figure}
\noindent {\bf Extended Data Fig.~5. Velocity dispersion profile for DLA0817g.} The observed velocity dispersion profile is measured from the standard deviation of a Gaussian fit to each pixel, and is shown by the data points where the data has been binned into bins equal to the size of the horizontal error bars. The vertical error bars reflect the 16 and 84 percentile spread in measurements per bin. The radius has been de-projected for the inclination of DLA0817g. The dashed line is the value derived from the kinematic modeling. The solid colored region is the 16 and 84 percentile spread in the constant velocity dispersion model, showing that the increase in velocity dispersion at the galactic center is due to beam-smearing. 

\newpage
\begin{figure}[!t]
\centering
\includegraphics[width=0.65\textwidth]{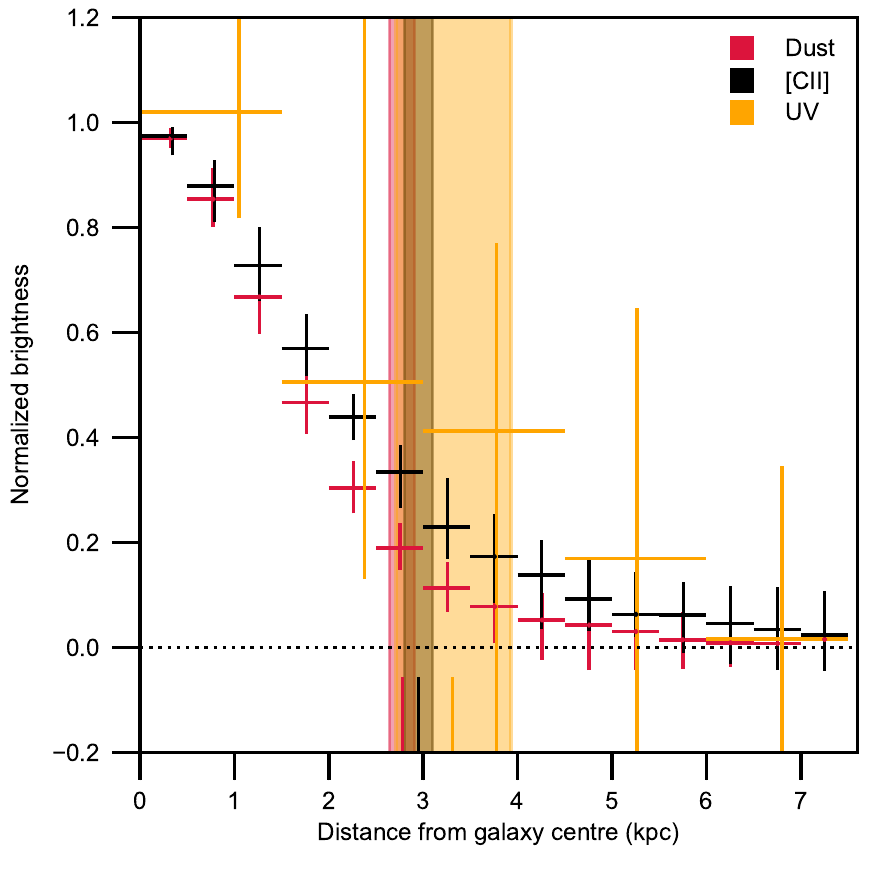}
\end{figure}
\noindent {\bf Extended Data Fig.~6. Light profile for dust continuum, \CII\ line and uv emission.} The light profiles are scaled by the emission at the kinematic center. Distances from the kinematic center are de-projected, taking into account the inclination of the disk. Both the dust continuum and the \CII\ emission are convolved with a Gaussian kernel to the slightly worse resolution of the UV observations. This increases the width of the surface density profile by $\approx$10\%. Uncertainties are estimated from the spread in measurements within the bins defined by the horizontal error bars. The \CII\ surface density profile, the dust continuum profile, and the UV surface density profile --within the uncertainties-- are consistent with each other. This can also be seen by the effective radii (marked by solid vertical lines), whose 1$\sigma$ uncertainties (marked by the vertical colored regions) overlap at a value of $\sim$3~kpc.

\newpage
\begin{figure}[!t]
\includegraphics[width=\textwidth]{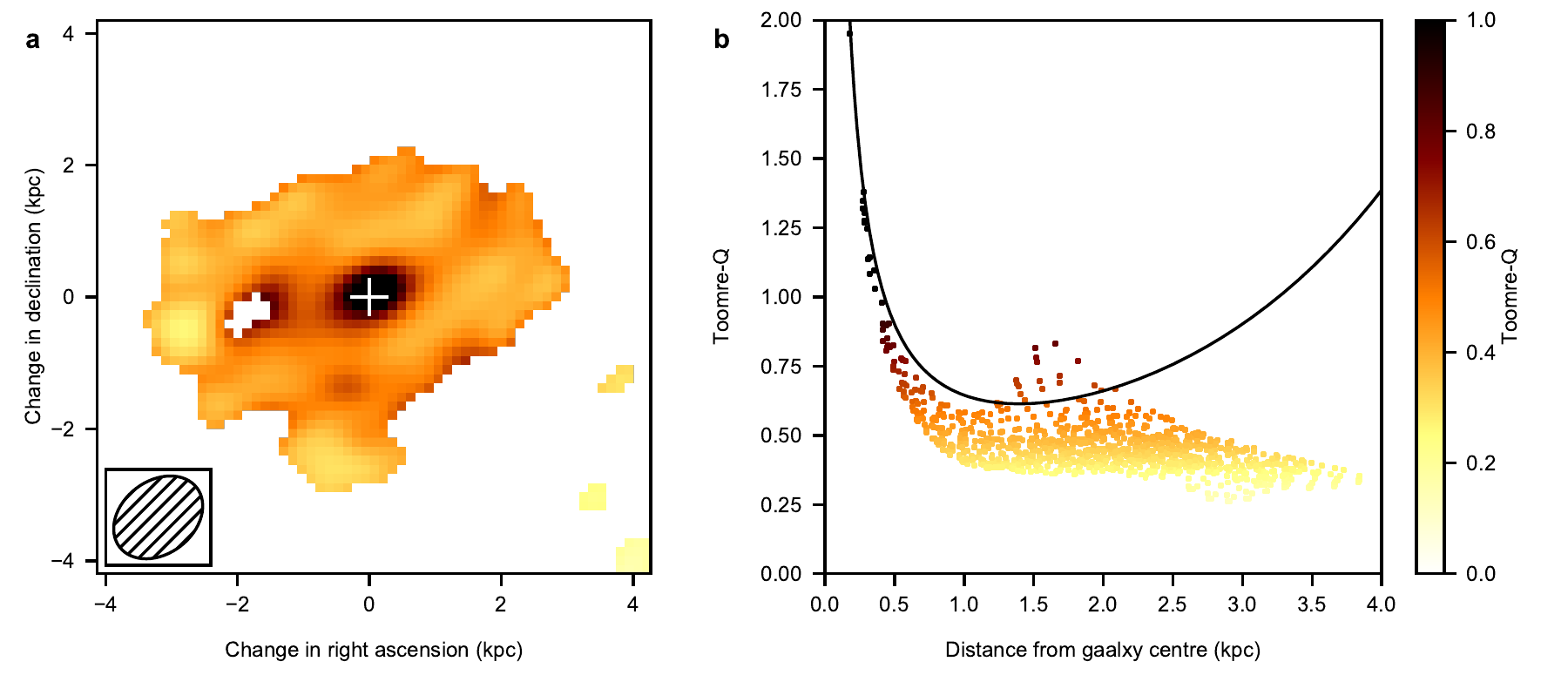}
\end{figure}
\noindent {\bf Extended Data Fig.~7. The Toomre-$\bm Q$ parameter for DLA0817g.} Panel A shows the spatial distribution of the Toomre-$Q$ parameter assuming that the gas density is traced by the \CII\ emission. This spatial distribution is still convolved with the ALMA synthesized beam. Over the entire disk, $Q$ is roughly constant and below 1, indicating that the disk is unstable against axisymmetric perturbations. The white cross shows the kinematic center of the emission and the inset shows the ALMA beam for the \CII\ emission. Panel~B shows the radial profile of the Toomre-$Q$ parameter. The solid dark line shows $Q$, assuming the gas density falls off exponentially, at the same rate as the \CII\ emission. The colored squares are the observed data, as in Panel~A, corrected for the projected radius. This panel shows that the observed data underestimates $Q$ at large radii due to beam smearing, which increases the emission, and thus the gas surface density.

\newpage
\begin{figure}[!t]
\centering
\includegraphics[width=0.8\textwidth]{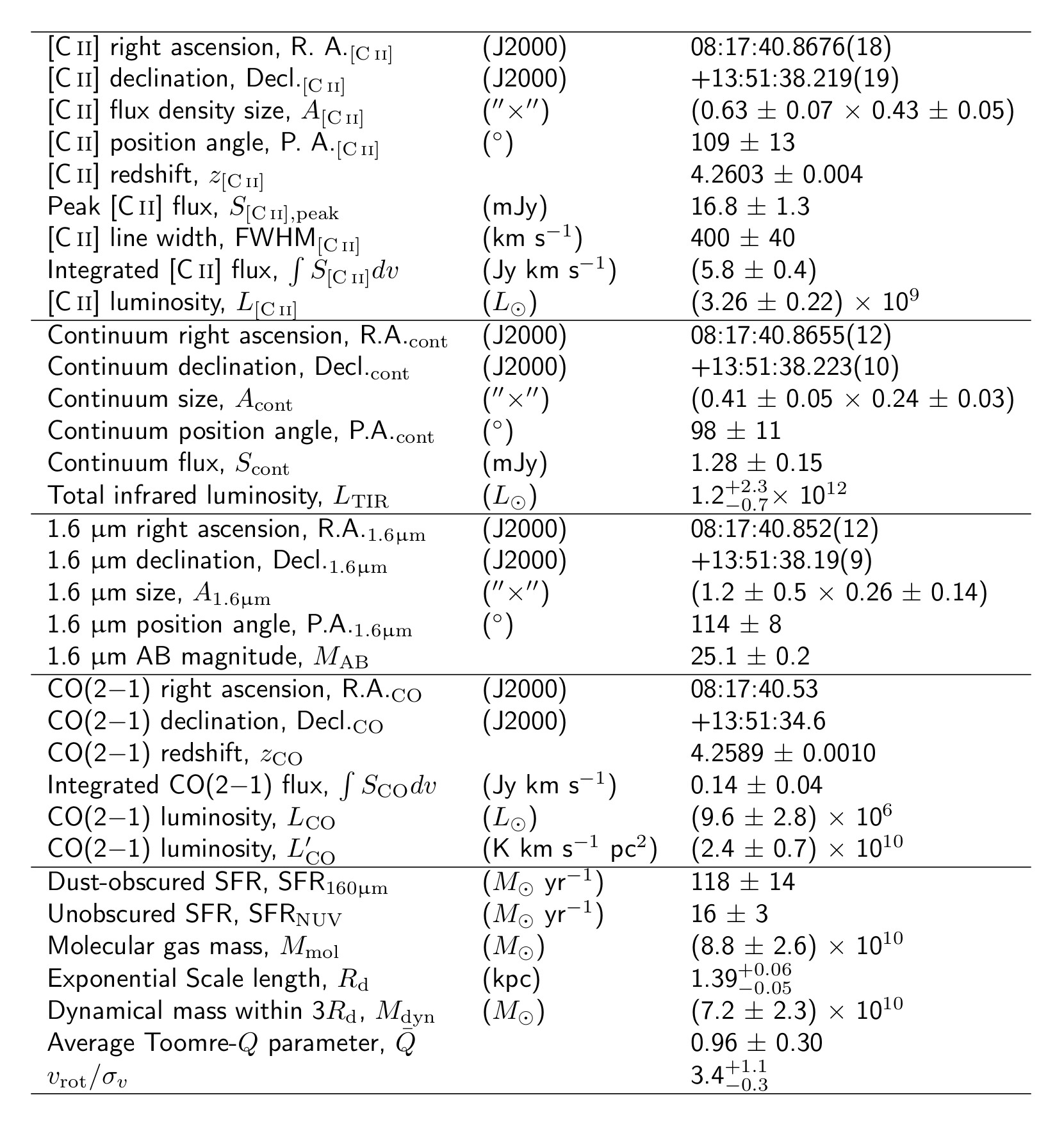}
\end{figure}
\noindent {\bf Extended Data Table~1. Physical Properties of DLA0817g.} Physical positions and sizes are estimated from two-dimensional Gaussian fits to the velocity-integrated \CII\ flux density and the far-infrared continuum emission, after deconvolving the ALMA synthesized beam. The redshift of the \CII\ emission and the full width at half maximum, FWHM$_{\rm [C\,{\scriptscriptstyle II}]}$, are derived from a double Gaussian fit to the integrated spectrum (Fig. S1). The quantities derived from the CO emission have been corrected for the effects of the cosmic microwave background.

\newpage
\begin{figure}[!t]
\centering
\includegraphics[width=0.8\textwidth]{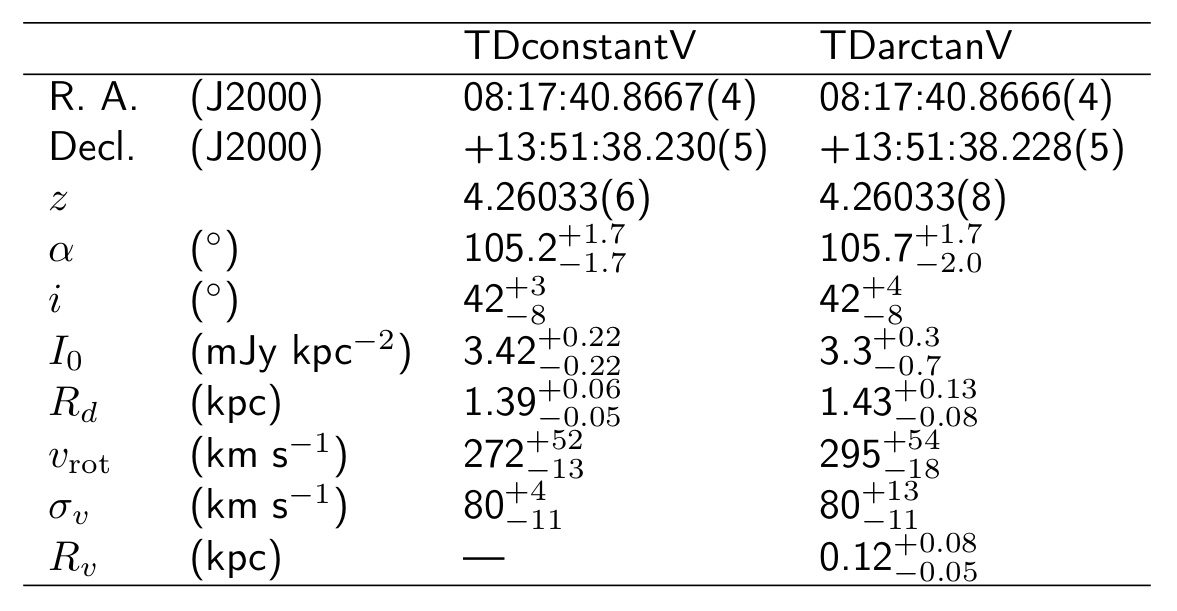}
\end{figure}
\noindent {\bf Extended Data Table~2. Results from the kinematic analysis of DLA0817g.} 
Two models are given, both assuming that the \CII\ emission arises from a thin disk, but one assuming a constant velocity profile (TDconstantV), and the other model, an arctangent velocity profile (TDarctanV). The central position of the kinematic center together with the position angle of the major axis, $\alpha$, and the inclination, $i$, determine the position and orientation of the disk. The intensity of the disk is determined by the central intensity, $I_0$, and the exponential scale radius, $R_{\rm d}$. Finally the kinematic information of the gas is contained within the maximum rotation velocity, $v_{\rm rot}$, and the dispersion of the gas $\sigma_v$. The arctangent velocity profile requires an additional scale radius, $R_v$. The number in parenthesis for the right ascension (R.A.), declination (Decl.) and redshift ($z$) represents the $1\sigma$ uncertainty in the last digit; for the remaining parameters, the uncertainties are asymmetric, and are hence listed as $1\sigma$ upper and lower bounds.

\end{document}